\documentclass[prd,aps,preprintnumbers,amsmath,amssymb,showpacs,showkeys,nofootinbib]{revtex4}
\usepackage{bm} 
\usepackage{graphicx}
\begin{document}
\title{Analytic Properties of Triangle Feynman Diagrams \\
in Quantum Field Theory} 
\author{Dmitri Melikhov%$^{a,b}$
}
\affiliation{
%$^a$ 
D.V.Skobeltsyn Institute of Nuclear Physics, \\
M.~V.~Lomonosov Moscow State University, 119991, Moscow, Russia\\
%$^b$ Faculty of Physics, University of Vienna, Boltzmanngasse 5, A-1090 Vienna, Austria
}
\date{\today}
\begin{abstract}
We discuss dispersion representations for the triangle diagram 
$F(p_1^2,p_2^2,q^2)$, the single dispersion 
representation in $q^2$ and the double dispersion representation in $p_1^2$ and $p_2^2$, with  
special emphasis on the appearance of the anomalous singularities and the anomalous
cuts in these representations. 

\end{abstract}
\maketitle
%%%%%%%%%%%%%%%%%%%%%%%%%%%%%%%%%%%%%%%%%%%%%%%%%%%%%%%%%%%
%%
%% BELOW IS THE CONTENT OF THE PAPER
%%
%%%%%%%%%%%%%%%%%%%%%%%%%%%%%%%%%%%%%%%%%%%%%%%%%%%%%%%%%%%%
%%%%%%%%%%%%%%%%%%%%%%%%%%%%%%%%%%%%%%%%%%%%%%%%%%%%%%%%%%%
%%
%% BELOW IS THE CONTENT OF THE PAPER
%%
%%%%%%%%%%%%%%%%%%%%%%%%%%%%%%%%%%%%%%%%%%%%%%%%%%%%%%%%%%%%
\section{Introduction}
Triangle diagrams have many applications in quantum field theory; let us recall some of such applications: 
they give the radiative corrections to the form factors of a relativistic 
particle, e.g., quark or electron; they describe the amplitudes of radiative and leptonic decays of
hadrons, e.g., $\pi^0\to\gamma\gamma$; they provide essential contributions to the amplitudes of
hadronic decays, such as $K\to 3\pi$; they give the main contribution to the weak and
electromagnetic form factors of relativistic bound states. Also, these diagrams are responsible 
for one of the most interesting phenomenon of quantum field theory --- for quantum anomalies. 

\begin{figure}[h]
\begin{center}
\includegraphics[width=7cm]{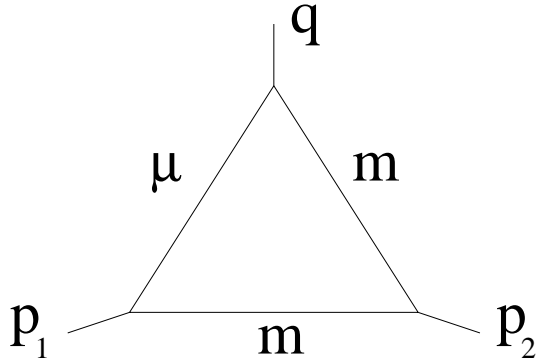} 
\caption{\label{fig:0.1} 
The Feynman diagram $F(p_1^2,p_2^2,q^2)$.}
\end{center}
\end{figure}

In this lecture, we discuss spectral representations for the one-loop triangle Feynman diagram 
with spinless particles in the loop (Fig.~\ref{fig:0.1}).\footnote{The inclusion of spin essentially does not
change the analysis and only leads to the technical but not conceptual complications.}
\begin{eqnarray}
\label{f}
F(q^2,p_1^2,p_2^2)=\frac{1}{(2\pi)^4 i}\int  
\frac{dk}{(m^2-k^2-i0)(\mu^2-(p_1-k)^2-i0)(m^2-(p_2-k)^2-i0)}, \nonumber\\
\qquad q=p_1-p_2.
\end{eqnarray}

The function $F$ is easily calculable in the Euclidean region of all spacelike external momenta 
but has complicated analytic properties in the Minkowski space relevant for
the description of processes with real particles. To handle these processes, dispersion
representations of the diagram are known to be very efficient. 

The application of the dispersion representations to the triangle diagram has a long history 
(see the original papers \cite{karplus,landau,fronsdal,norton} and the corresponding chapters 
in handbooks and reviews, e.g., \cite{burton,anisovich_book,landau_lifshitz,iz,zwicky}). 
This lecture is based to a large extent on the material of our paper \cite{lms}. 

One can consider single (in the variable $q^2$) and double (in the variables $p_1^2$ and $p_2^2$) dispersion representations 
for the triangle diagram. 

An essential feature of the single spectral representation is the appearance of the 
anomalous threshold \cite{karplus} and the anomalous contribution to the spectral representation 
in a specific region of the variables $p_1^2$, $p_2^2$, $\mu$, and $m$: 
this {\it anomalous} threshold in $q^2$ is located below the {\it normal}, or unitary, threshold, related to the 
possible physical intermediate states in the unitarity relation. As a result, it is the anomalous singularity that mainly  
determines the properties of the triangle diagrams in the region of small $q^2$. The location of the anomalous singularity and the 
anomalous threshold in the single spectral representation is obtained by solving the Landau equations \cite{landau,iz,hagop}. 

The double spectral representation in $p_1^2$ and $p_2^2$ for the case of the decay kinematics 
$0<q^2<(\mu-m)^2$ also has an anomalous contribution; the latter is, however, of a different 
kind than the one in the single representation in $q^2$. In principle, all anomalous contributions 
are related to the motion of a branch point of the integrand from the unphysical sheet 
onto the physical sheet through the normal cut and the corresponding modification of the integration contour 
in the complex plane of the appropriate variable. However, the location of the anomalous threshold in the double spectral 
representation in the decay region $0<q^2<(\mu-m)^2$ is not determined by the Landau rules \cite{braun}. 
The anomalous threshold in the double spectral representation lies beyond the normal
threshold, and the anomalous piece dominates the double dispersion representation for the triangle diagram in the region $q^2\simeq(\mu-m)^2$. 

An exhaustive analysis of the single and the double dispersion 
representations of the triangle diagram for all
values of the external and the internal masses can be found in \cite{fronsdal}. 

We discuss here the single and the double dispersion representations of the triangle diagram, 
with the emphasis on the properties of the anomalous contributions. We point out that in many 
cases the application of the double spectral 
representation in $p_1^2$ and $p_2^2$ is technically much simpler than the application of the 
single representation in $q^2$.

We start, in Section \ref{sect:ii}, with the case of particles of the same mass in the loop. 
We illustrate the appearance of the anomalous cut in the single spectral 
representation in $q^2$ 
for $p_1^2>0$, $p_2^2>0$, and $p_1^2+p_2^2\ge 4m^2$. This spectral representation 
has a rather complicated form especially for complex values of $p_1^2$ and $p_2^2$. 

In Section \ref{sect:iii}, we then discuss the double spectral representation in $p_1^2$ and
$p_2^2$. This representation is very simple for $q^2<0$ and contains only the normal cut. 
This makes the 
application of the double spectral representation particularly convenient for the 
analysis of processes 
described by the triangle diagram for timelike $p_1$ and $p_2$ 
in the region $p_1^2+p_2^2\ge 4m^2$ and for higher overthreshold values of $p_1^2$ and $p_2^2$. 

In Section \ref{sect:iv}, we discuss the double spectral representation in 
$p_1^2$ and $p_2^2$ for the case of particles of different masses in the loop. 
Here, in the case of the decay kinematics $0<q^2<(\mu-m)^2$ , the anomalous 
contribution to the double spectral representation emerges. 
We emphasize that the double spectral representation in $p_1^2$ and $p_2^2$ provides a very convenient 
tool for considering processes at overthreshold values of the variables $p_1^2$ and $p_2^2$, 
relevant for the decay processes, such as, e.g., $K\to 3\pi$ decays. 

%\newpage
\section{\label{sect:ii}Spacelike momentum transfers, equal masses in the loop}
In this Section, we consider the case of particles of the same mass $m$ 
in the loop and $q^2<0$, but do not restrict the values of $p_1^2$ and $p_2^2$.  

\subsection{Single dispersion representation in $q^2$}
A normal single dispersion representation in $q^2$ may be written as 
\begin{eqnarray}
\label{single}
F(q^2,p_1^2,p_2^2)=\frac{1}{\pi}\int \frac{dt}{t-q^2-i0}\sigma(t,p_1^2,p_2^2).    
\end{eqnarray} 
For $p_1^2<0$ and $p_2^2<0$, the absorptive part $\sigma(t,p_1^2,p_2^2)$ may be calculated by the 
Cutkosky rules, i.e., by placing particles attached to the $q^2$ vertex on the mass shell  
$(m^2-k^2-i0)^{-1}\to 2i\pi\theta(k_0)\delta(m^2-k^2)$. 
The result reads 
\begin{eqnarray}
\sigma(t,p_1^2,p_2^2)&=&\frac{1}{16\pi\lambda^{1/2}(t,p_1^2,p_2^2)}
\log\left(
\frac{t-p_1^2-p_2^2+\lambda^{1/2}(t,p_1^2,p_2^2)\sqrt{1-4m^2/t}}
{t-p_1^2-p_2^2-\lambda^{1/2}(t,p_1^2,p_2^2)\sqrt{1-4m^2/t}}\right)\theta(t-4m^2). 
\end{eqnarray}
The function $\sigma(t,p_1^2,p_2^2)$ has the branch point of the logarithm at 
$q^2=t_0(p_1^2,p_2^2)$ given by the solution to the equation 
$(t-p_1^2-p_2^2)^2=\lambda(t,p_1^2,p_2^2)(1-4m^2/t)$, or,
equivalently, to the equation  
\begin{eqnarray}
\frac{p_1^2p_2^2t}{m^2}+\lambda(p_1^2,p_2^2,t)=0.  
\end{eqnarray} 
Explicitly, one finds \cite{karplus,landau}
\begin{eqnarray}
\label{t0}
t_0^{\pm}(p_1^2,p_2^2)=p_1^2+p_2^2-\frac{p_1^2p_2^2}{2m^2}\pm
\frac{1}{2m^2}R(p_1^2)R(p_2^2),
\end{eqnarray}
where the function $R(p^2)$ at $p^2<0$ reads
\begin{eqnarray}
\label{R1}R(p^2)=\sqrt{p^2(p^2-4m^2)},\qquad p^2<0.
\end{eqnarray}
To obtain $R(p_i^2)$ ($i=1,2$) at $p_i^2>0$, one substitutes \cite{landau_lifshitz}
\begin{eqnarray}
\label{xi}p_i^2=-m^2\frac{(1-\xi_i)^2}{\xi_i}, \qquad i=1,2.
\end{eqnarray}
%\newpage
%\newpage 
This transformation maps the upper half-plane of the complex variable $p_i^2$ onto the 
internal semicircle with unit radius in the complex $\xi$-plane: 
the region $0<\xi_i<1$ corresponds to $p_i^2<0$, 
the boundary of the semicircle  $\xi_i=\exp(i \varphi_i)$, $0< \varphi_i <\pi$,  
corresponds to the unphysical region $0<p_i^2<4m^2$, 
and the segment $-1<\xi_i<0$ corresponds to $4m^2<p_i^2$. Then 
\begin{eqnarray}
R(p_i^2)=m^2\frac{1-\xi_i^2}{\xi_i}, 
\end{eqnarray} 
and, for $0<p_i^2<4m^2$, we obtain 
\begin{eqnarray}
R(p_i^2)=-2i \sin \varphi_i. 
\end{eqnarray} 
Finally, $R(p^2)$, as obtained by analytic continuation, reads
\begin{eqnarray}
\label{R}R(p^2)=\left\{\begin{array}{ll}
\sqrt{p^2(p^2-4m^2)}&\qquad\mbox{for}\ p^2<0,\\[1ex]
-i\sqrt{p^2(4m^2-p^2)}&\qquad\mbox{for}\ 0<p^2<4m^2,\\[1ex]
-\sqrt{p^2(p^2-4m^2)}&\qquad\mbox{for}\ 4m^2<p^2.
\end{array}\right.
\end{eqnarray}
Note the sign of $R(p^2)$ for $p^2>4m^2$, which signals that the
square-root function is now on its negative branch.

With the definition of $t_0^{\pm}(p_1^2,p_2^2)$ given by
(\ref{t0}) and (\ref{R}), one can study the trajectories of $t_0^{\pm}(p_1^2,p_2^2)$. 

\subsection{The branch point $t_0^+(p_1^2,p_2^2)$}
\begin{figure}[h!]
\begin{center}
\begin{tabular}{c}
\includegraphics[width=7.5cm]{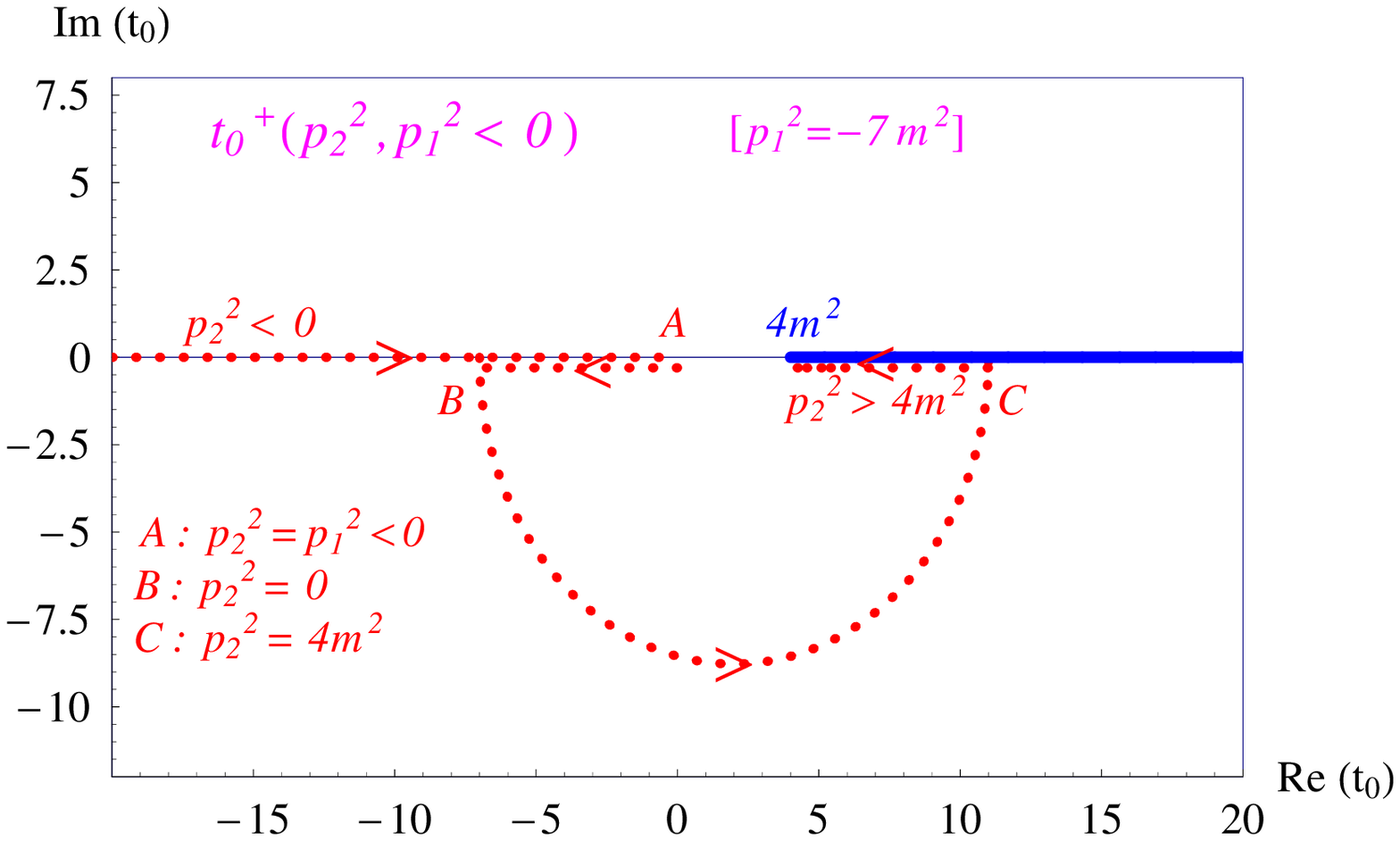}
\includegraphics[width=7.5cm]{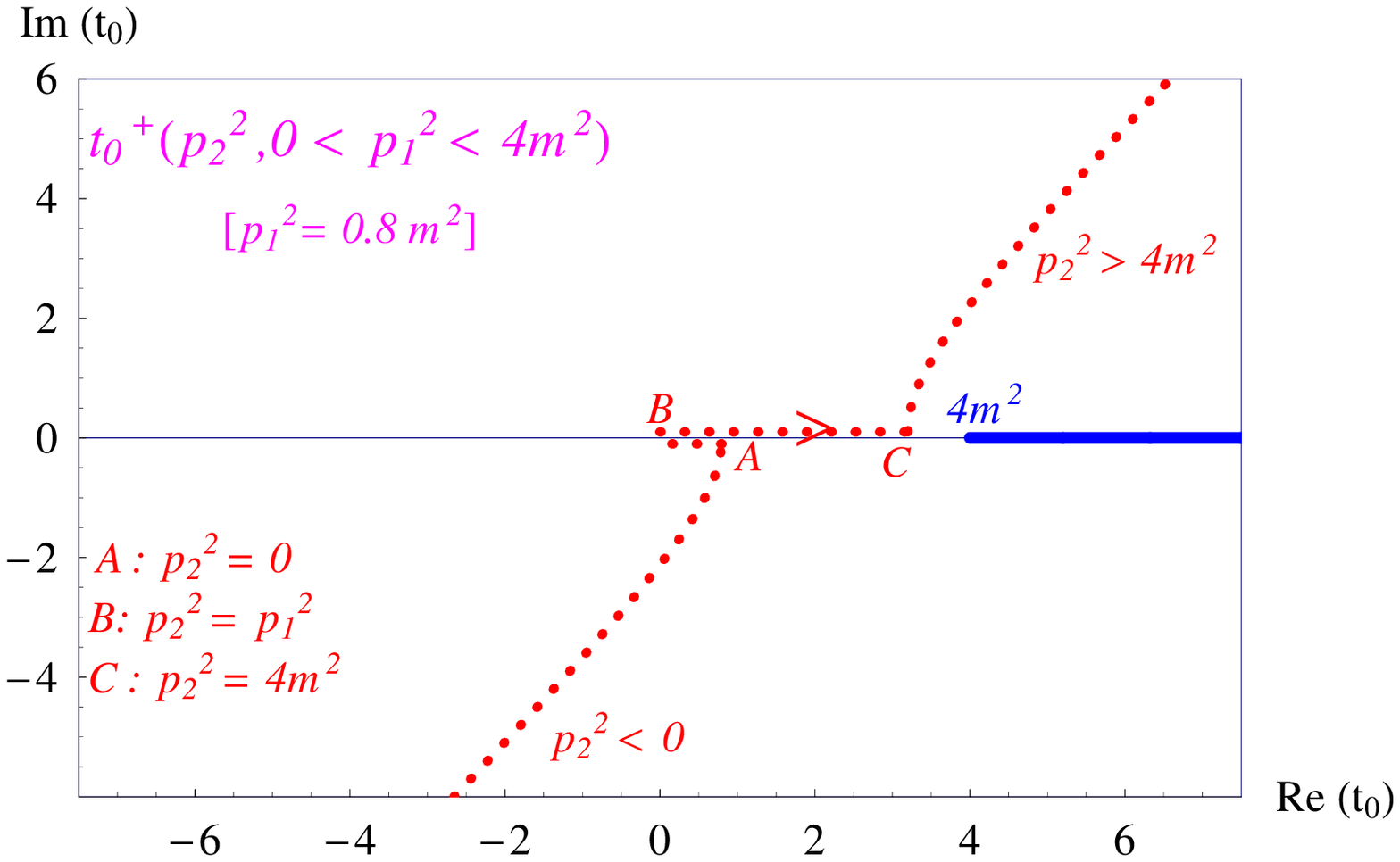}\\
\qquad \qquad (a) \hspace{7cm}  (b)           \\
\\
\includegraphics[width=7.5cm]{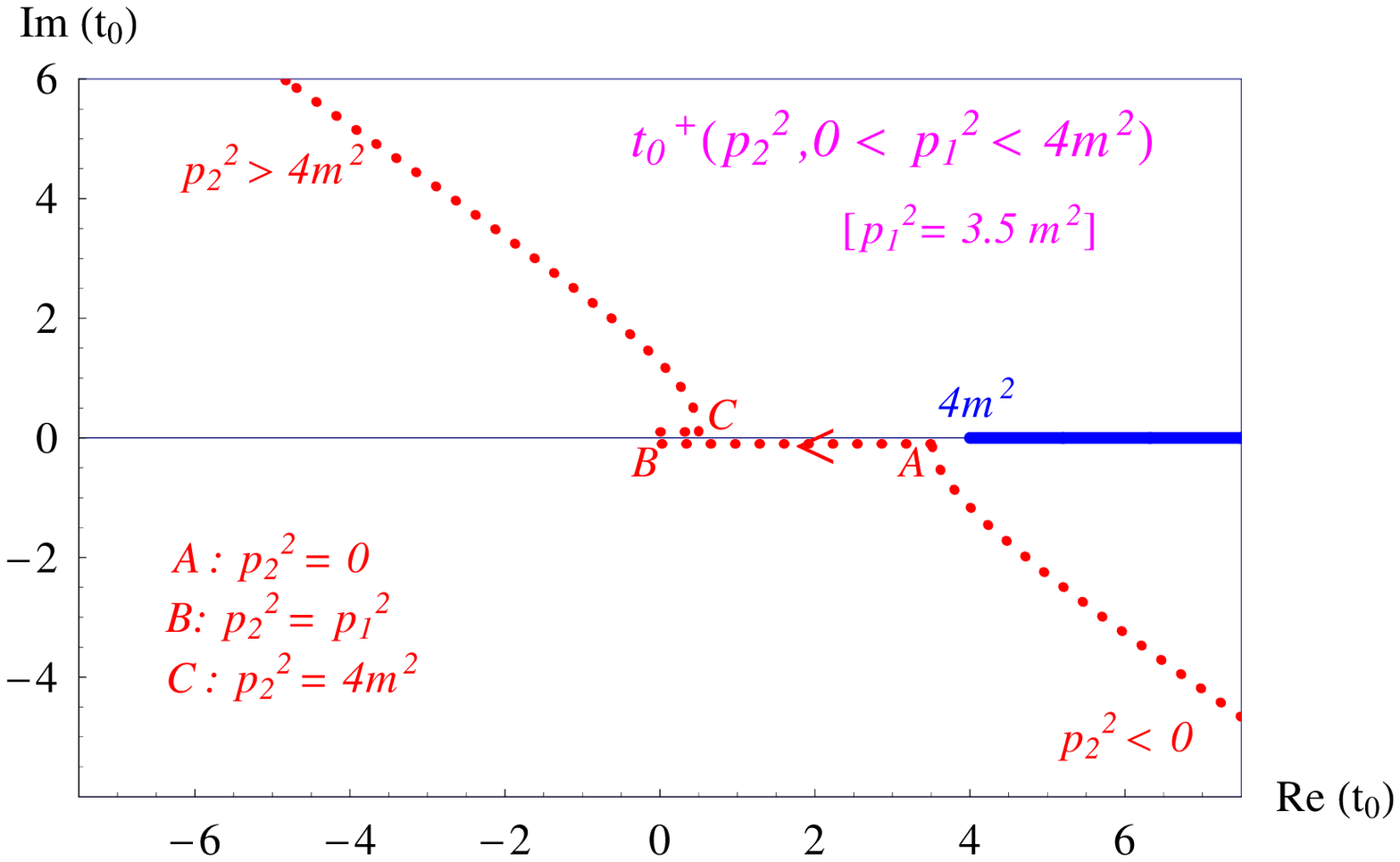}
\includegraphics[width=7.5cm]{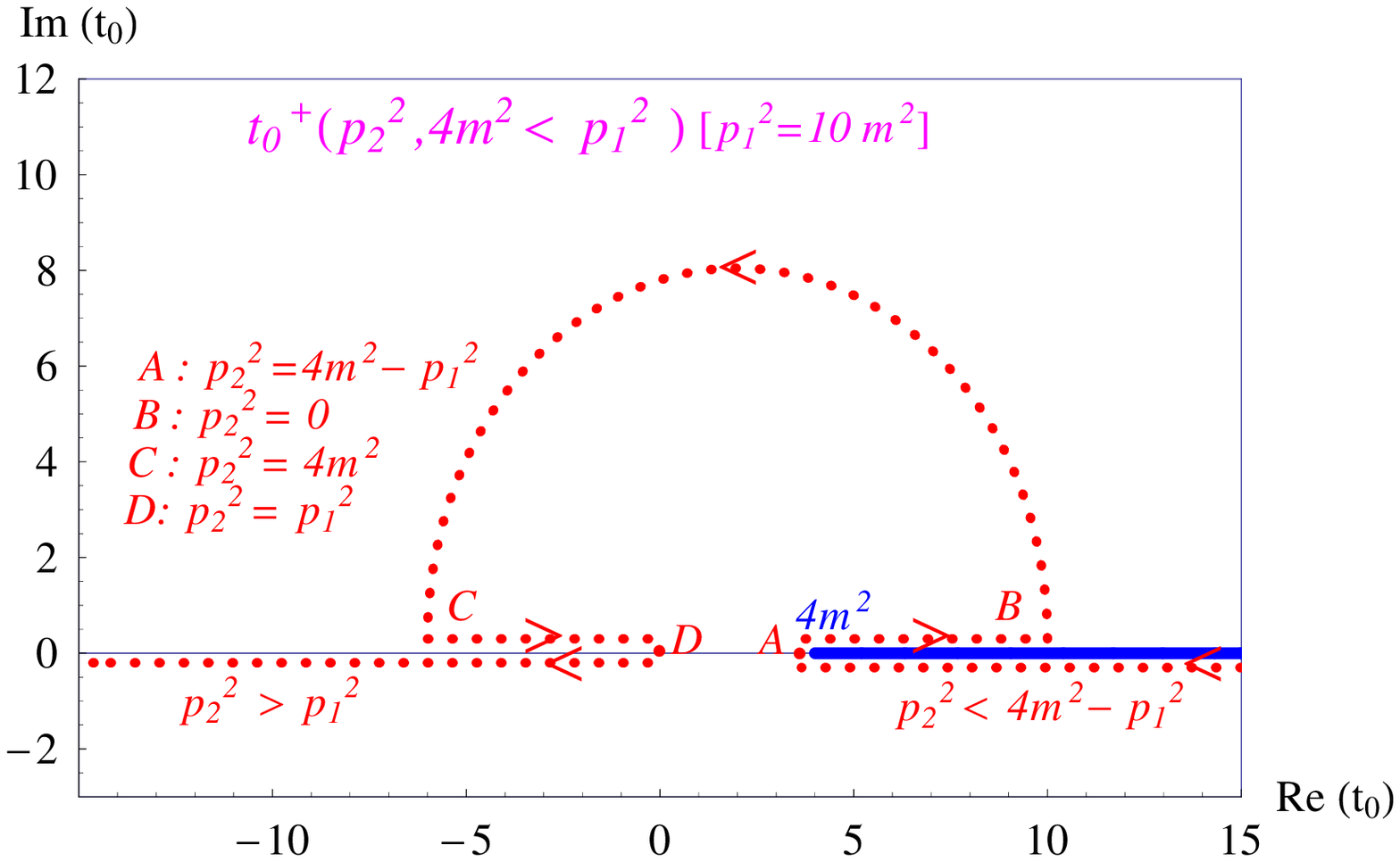}\\
\qquad \qquad (c) \hspace{7cm}  (d)           \\
\\
\end{tabular}
\caption{\label{fig:4}
The trajectory of the branch point
$t^+_0(p_1^2,p_2^2)$ vs $p_2^2$ at a fixed value of $p_1^2$ in the
complex-$q^2$ plane. Dashed lines denote trajectories on the second (unphysical) sheet. 
(a) $p_1^2\le0$ [$p_1^2=-7 m^2$].  
(b,c) $0<p_1^2<4m^2$: (b) $p_1^2=0.8 m^2$; (c) $p_1^2=3.5 m^2$. 
(d) $4m^2 < p_1^2$ [$p_1^2=10 m^2$]. The normal cut along the real axis for $q^2>4m^2$ is depicted in~blue.
Noteworthy, the branch point $t^+_0(p_1^2,p_2^2)$ always remains on the second (unphysical) sheet of the Riemann surface and never 
migrates onto the physical sheet. Therefore, $t^+_0(p_1^2,p_2^2)$ does not influence the location of the cut on the physical sheet.
}
\end{center}
\end{figure}

The branch point $t_0^+(p_1^2,p_2^2)$, treated as the function of $p_2^2$ at fixed $p_1^2$, 
has a rather cumbersome trajectory on the second sheet of the complex plane, but never appears~on~the
physical sheet (see Fig.\ref{fig:4}). This means that the motion of this branch point does not influence 
the single $q^2$-spectral representation for $F(q^2,p_1^2,p_2^2)$.
%\newpage

%\newpage FIGURE 3 width 8.2 cm
\begin{figure}[!t]
\begin{center}
\begin{tabular}{c}
\includegraphics[width=7.5cm]{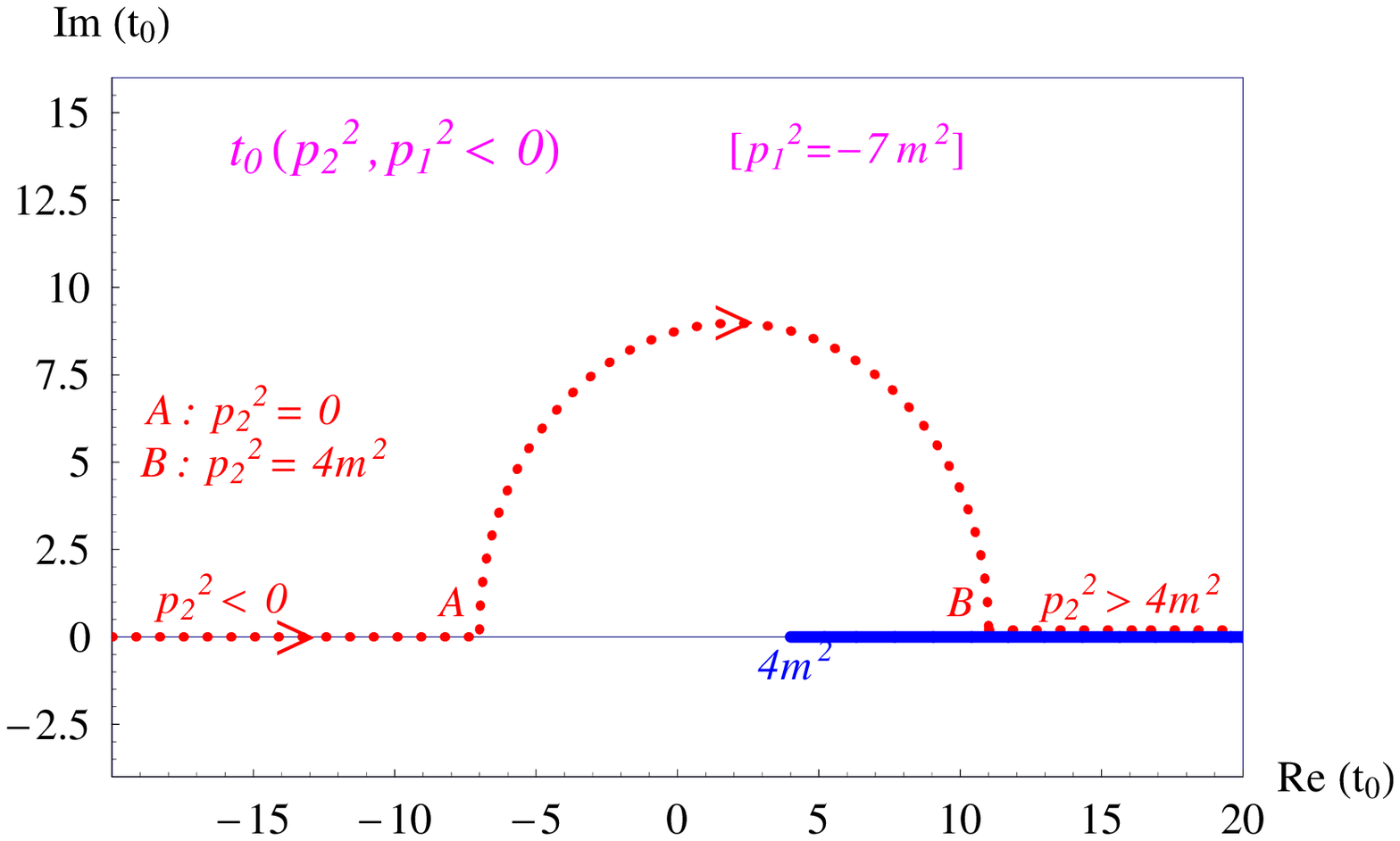}
\includegraphics[width=7.5cm]{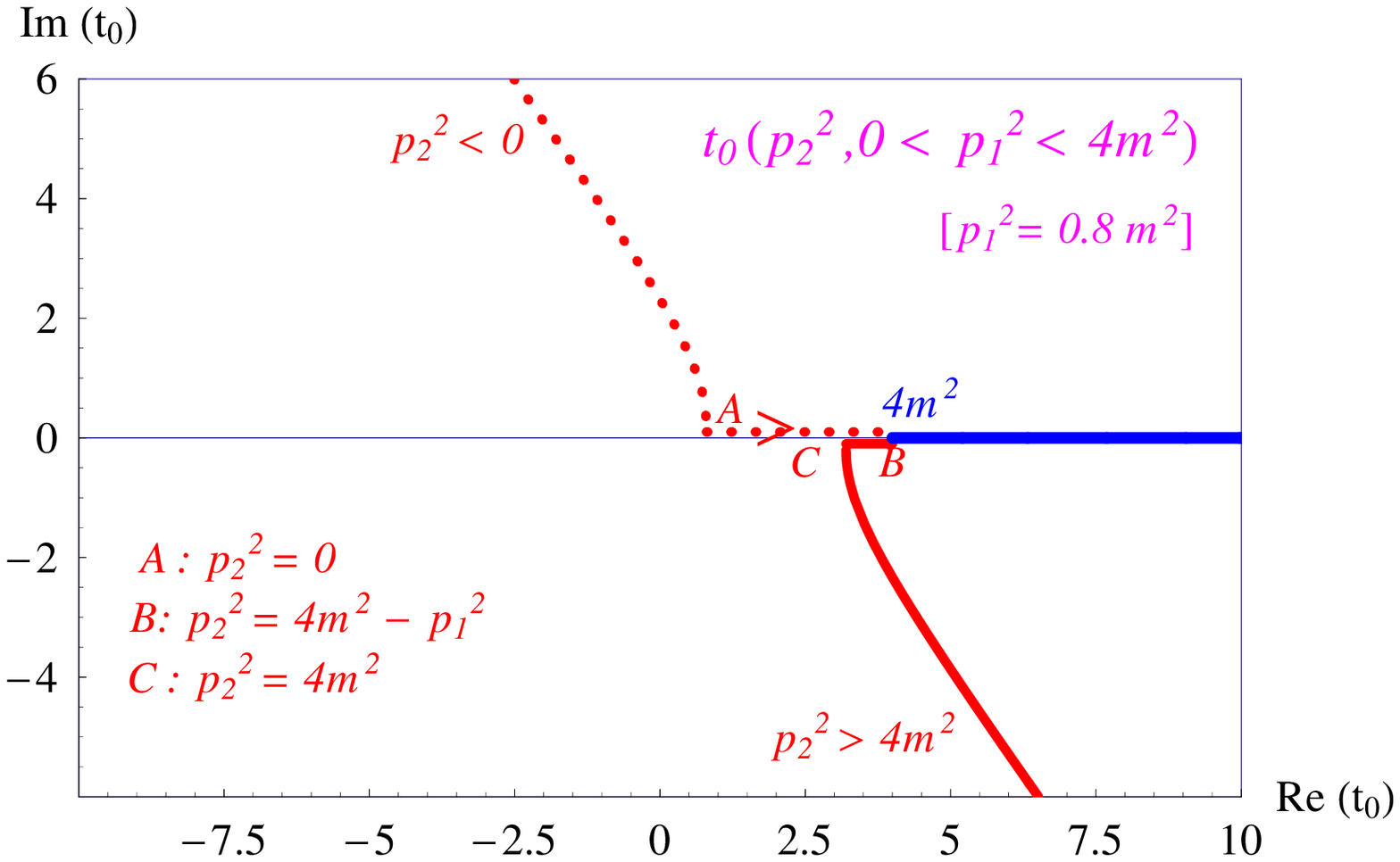}\\
\qquad \qquad (a) \hspace{7cm}  (b)           \\
\\
\includegraphics[width=7.5cm]{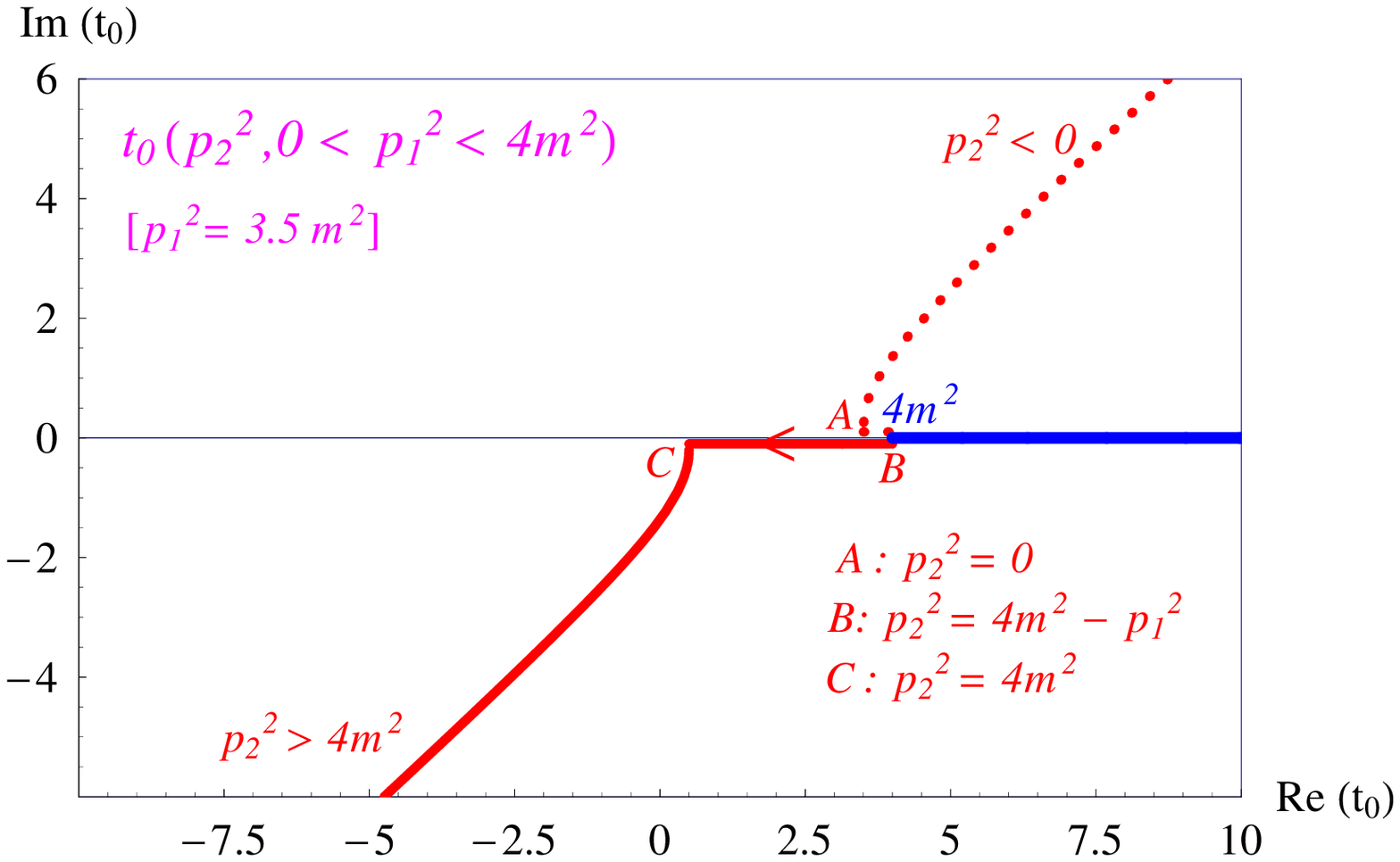}
\includegraphics[width=7.5cm]{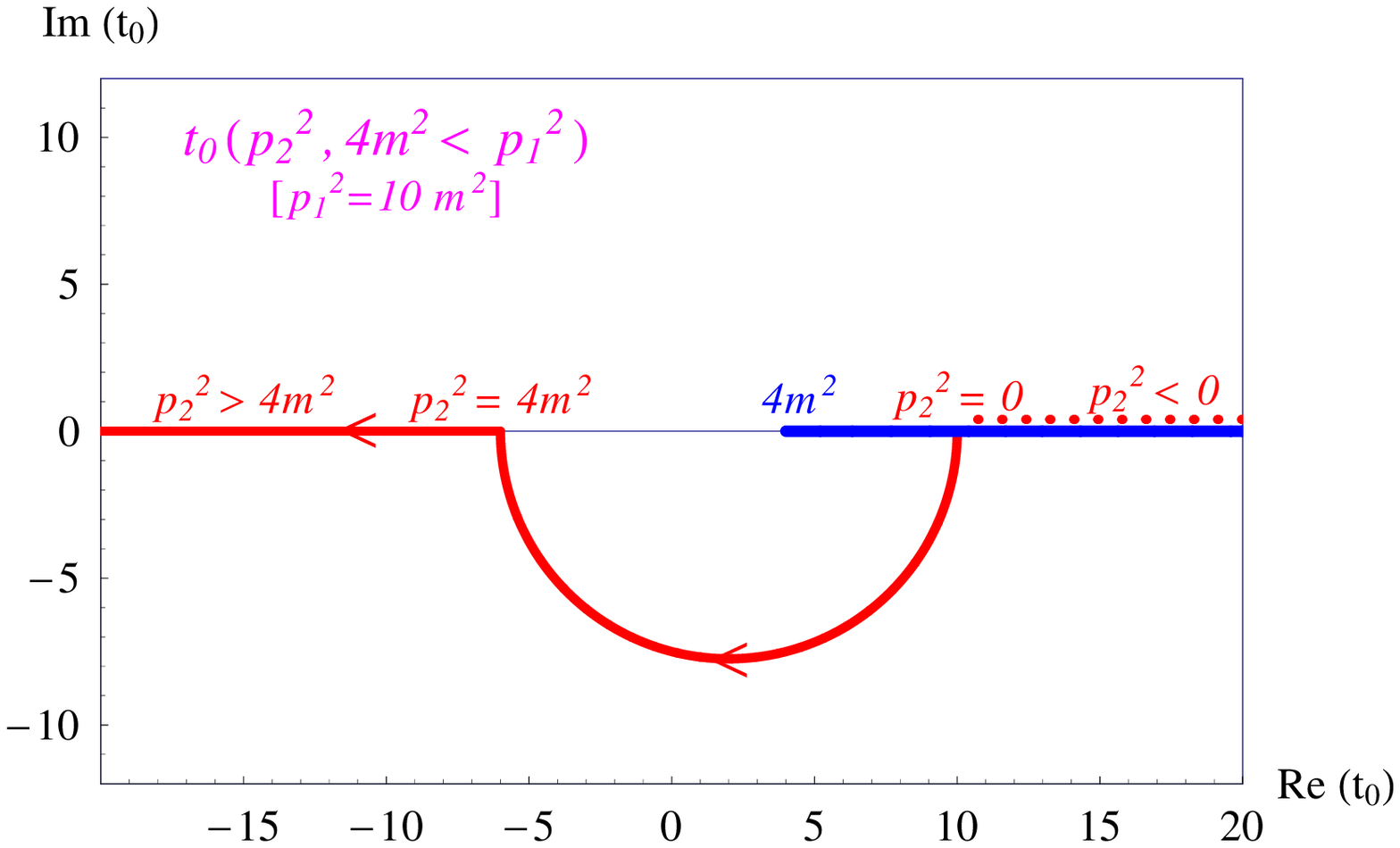}\\
\qquad \qquad (c) \hspace{7cm}  (d)           \\
\\
\end{tabular}
\caption{\label{fig:3}
The trajectory of the branch point
$t_0(p_1^2,p_2^2)\equiv t^-_0(p_1^2,p_2^2)$ vs $p_2^2$ at a fixed value of $p_1^2$ in the
complex-$q^2$ plane: (a) $p_1^2\le0$ [$p_1^2=-7m^2$]: the (dashed~line) trajectory
lies on the second sheet, and does not appear on the physical
sheet; (b,c) $0<p_1^2<4m^2$ [(b): $p_1^2=0.8m^2$, (c): $p_1^2=3.5m^2$]: 
the trajectory first lies~on~the second sheet (dashed line), but for $p_2^2 > 4m^2-p_1^2$ moves around the
normal-cut branch point through the normal cut onto the physical
sheet (solid line); (d) $4m^2 < p_1^2$ [$p_1^2=10m^2$]: for $p_2^2<0$, the trajectory runs on the unphysical sheet, but at $p_2^2=0$ it comes up to
the physical sheet and for $p_2^2>0$ travels over this physical sheet. 
As soon as the branch point $t_0(p_1^2,p_2^2)$ appears on the physical sheet, the $q^2$-integration contour should be modified such that it embraces 
both the normal cut and the branch point $t_0(p_1^2,p_2^2)$. The normal cut along the real axis for $q^2>4m^2$ is depicted in~blue.
}
\end{center}
\end{figure}
\subsection{The branch point $t_0^-(p_1^2,p_2^2)$}
Treated as the function of $p_2^2$ for a fixed value of $p_1^2$, the branch point 
$t_0\equiv t_0^-(p_1^2,p_2^2)$ has its trajectory at $p_2^2<0$ on the unphysical sheet 
of the complex $q^2$-plane. 
However, at some value of $p_2^2$, it crosses the cut at $t\ge 4m^2$. Crossing the cut means that the branch point 
$t_0$ moves onto the physical sheet through this cut. The integration contour in the Cauchi theorem is 
chosen on the physical sheet of complex variable $q^2$ and embraces (any) region of complex variable $q^2$, 
where the function $F(q^2,p_1^2,p_2^2)$ of the complex variable $q^2$ has no singularities. 
The migration of the branch point $t_0$ onto the physical sheet pushes the integration contour away from the normal threshold 
and leads to the 
appearance of a new segment of the $q^2$-integration on the physical sheet \cite{lms,procura}. 
This new segment is called the anomalous cut. 
The trajectories of $t_0(p_1^2,p_2^2)$ vs $p_2^2$ for four different fixed values of $p_1^2$ are shown in Fig.~\ref{fig:3}. 
The dotted lines denote the segments of the trajectory lying on the 
unphysical sheet; the solid lines denote those segments that are on the physical sheet.

Therefore, for external momenta satisfying the relation 
$p_1^2>0$, $p_2^2>0$, $p_1^2+p_2^2>4m^2$, the integration contour in the dispersion representation 
for the form factor depends on the values of $p_1^2$ and $p_2^2$: the contour should be chosen 
such that it embraces both branch points: the normal branch point at $q^2=4m^2$ and the anomalous 
branch point at $q^2=t_0(p_1^2,p_2^2)$. 

\begin{figure}[!h]
\begin{center}
\includegraphics[width=14cm]{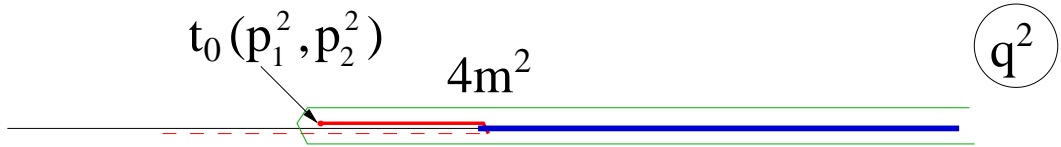}
\caption{\label{fig:new4} 
The integration contour (green) in the complex $q^2$-plane for 
$0<p_1^2<4m^2$, $0<p_2^2<4m^2$, and $p_1^2+p_2^2>4m^2$: it embraces the anomalous cut (red), which 
lies along the real axis from $t_0$ to $4m^2$, and the normal cut (blue) from $4m^2$ to $+\infty$.}
\end{center}
\end{figure}

Let us consider the single dispersion representation for the form factor in the region 
$0< p_1^2 < 4m^2$, $0 < p_2^2 < 4m^2$, and $4m^2 < p_1^2 +p_2^2$.
This case corresponds to an interesting example of a two-particle bound state, 
and is necessary for considering the nonrelativistic expansion.  
The corresponding $t_0$-trajectory is shown in Fig.~\ref{fig:3}. 
Fig.~\ref{fig:new4}  gives the integration contour for this case: this contour may be chosen along the real axis from 
$t_0(p_2^2)$ to $+\infty$. It contains two pieces: the normal part from $4m^2$ to $+\infty$
and the anomalous part from $t_0$ to $4m^2$. 

Let us start with the normal part, which has the form  
\begin{eqnarray}
\label{norm}
\sigma_{\rm norm}(t,p_1^2,p_2^2)=
\left\{\begin{array}{ll} 
\frac1{16\pi\sqrt{\lambda(t,p_1^2,p_2^2)}}
\log\left(
\frac{t-p_1^2-p_2^2+\lambda^{1/2}(t,p_1^2,p_2^2)\sqrt{1-4m^2/t}}
{t-p_1^2-p_2^2-\lambda^{1/2}(t,p_1^2,p_2^2)\sqrt{1-4m^2/t}}\right), 
\quad
(\sqrt{p_1^2}+\sqrt{p_2^2})^2\le t,\\ 
\frac1{8\pi\sqrt{-\lambda(t,p_1^2,p_2^2)}}\arctan
\left(\frac{\sqrt{-\lambda(t,p_1^2,p_2^2)}\sqrt{1-4m^2/t}}{t-p_1^2-p_2^2}\right), 
\quad  
p_1^2+p_2^2\le t\le (\sqrt{p_1^2}+\sqrt{p_2^2})^2,\\ 
\frac1{8\pi\sqrt{-\lambda(t,p_1^2,p_2^2)}}
\left[\pi+\arctan
\left(\frac{\sqrt{-\lambda(t,p_1^2,p_2^2)}\sqrt{1-4m^2/t}}{t-p_1^2-p_2^2}\right)\right], 
\quad  4m^2\le t\le p_1^2+p_2^2.
\end{array}\right. \nonumber
\end{eqnarray}
%\vspace{-1.6cm}
\vspace{-1cm}
\begin{eqnarray}
{}
\end{eqnarray}
Notice that the normal spectral density does not vanish at the normal threshold $t=4m^2$. 

The discontinuity of the form factor $F(q^2,p_1^2,p_2^2)$ on the anomalous cut is related 
to the discontinuity of the function $\sigma_{\rm norm}(t,p_1^2,p_2^2)$ and reads 
\begin{eqnarray}
\label{anom}
\sigma_{\rm anom}(t,p_1^2,p_2^2)=
\frac{1}{8\sqrt{-\lambda(t,p_1^2,p_2^2)}},\qquad t_0\le t\le 4m^2.
\end{eqnarray} 
Therefore, the full spectral density has the form 
\begin{eqnarray}
\sigma(t,p_1^2,p_2^2)=
\theta(p_1^2+p_2^2-4m^2)
\theta(t_0\le t\le 4m^2)\sigma_{\rm anom}(t,p_1^2,p_2^2)
+
\theta(4m^2\le t)\sigma_{\rm norm}(t,p_1^2,p_2^2). 
\end{eqnarray}
Clearly, the spectral density given by Eqs. (\ref{norm}) and (\ref{anom}) is a continuous function for 
$t>t_0$.
The spectral representation for the form factor then contains the normal and the anomalous contributions:   
\begin{eqnarray}
F(q^2,p_1^2,p_2^2)=
\theta(p_1^2+p_2^2-4m^2)
\int\limits_{t_0(p_1^2,p_2^2)}^{4m^2}
\frac{dt}{\pi(t-q^2-i0)}\sigma_{\rm anom}(t,p_1^2,p_2^2)%\nonumber\\
+
\int\limits_{4m^2}^{\infty}
\frac{dt}{\pi(t-q^2-i0)}\sigma_{\rm norm}(t,p_1^2,p_2^2). 
\end{eqnarray}
For $t_0(p_1^2,p_2^2)<q^2<4m^2$ (in case $p_1^2+p_2^2>4m^2$)
the imaginary part of the form factor comes from the anomalous
part, while for $q^2>4m^2$ it comes from the normal part. 

%\newpage

%*********************************************************************************
%\newpage
\subsection{An illustration for the case of equal external masses $p_1^2=p_2^2$}
Let us now specify the general formulas given above for the case of the equal ``external'' masses. 
We set $p_1^2=p_2^2=M^2$, and aim at obtaining the dispersion representation for the case $2m^2<M^2<4m^2$. 
Notice that both $p_1^2$ and $p_2^2$ are {\it below the normal thresholds}; the latter are located at $p_1^2=4m^2$ and $p_2^2=4m^2$.
Nevertheless, we shall see that the anomalous threshold in the single dispersion representation in $q^2$ will emerge.  

\begin{figure}[!ht]
\begin{center}
\includegraphics[width=10.9cm]{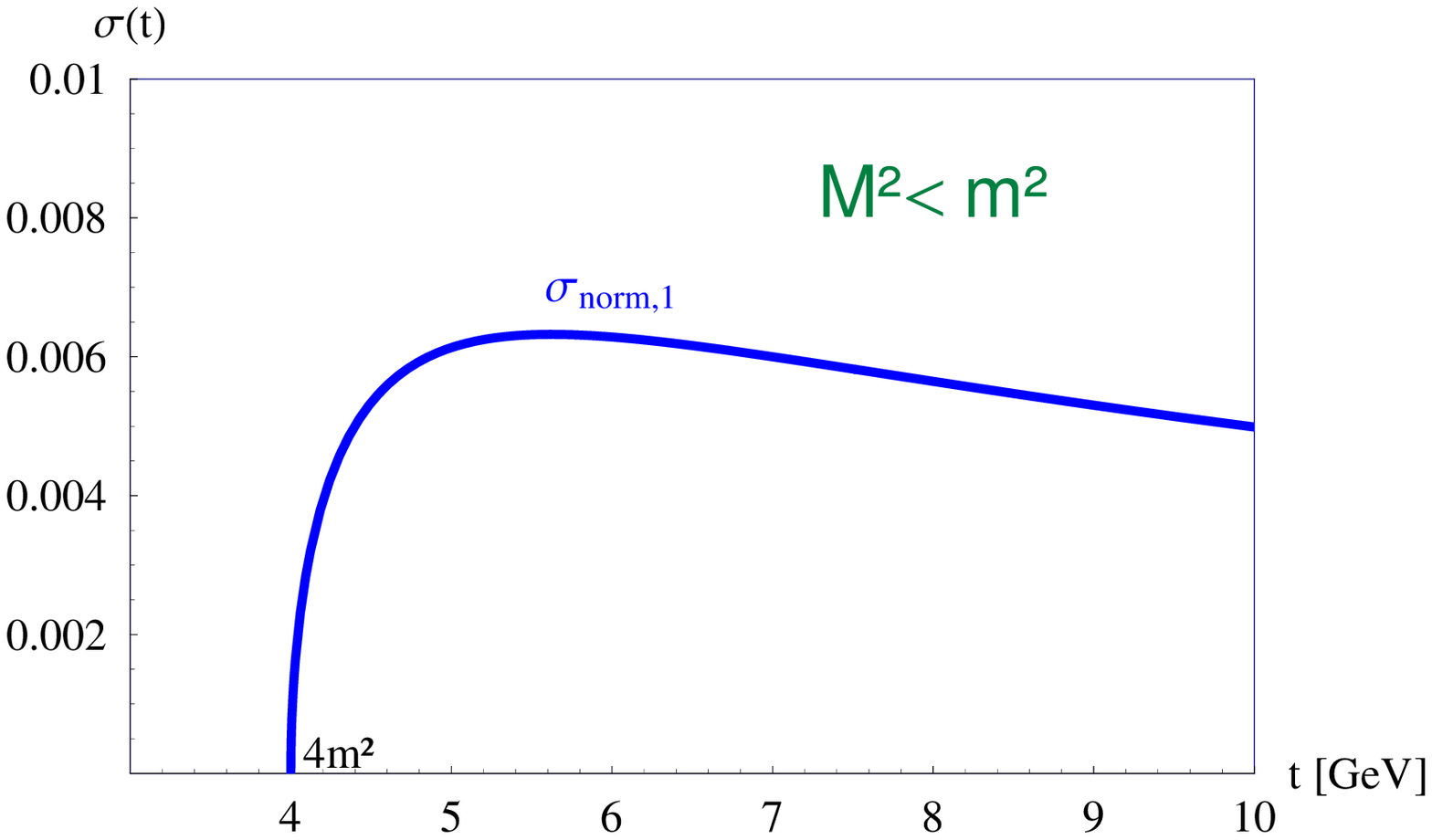}\\
\includegraphics[width=10.9cm]{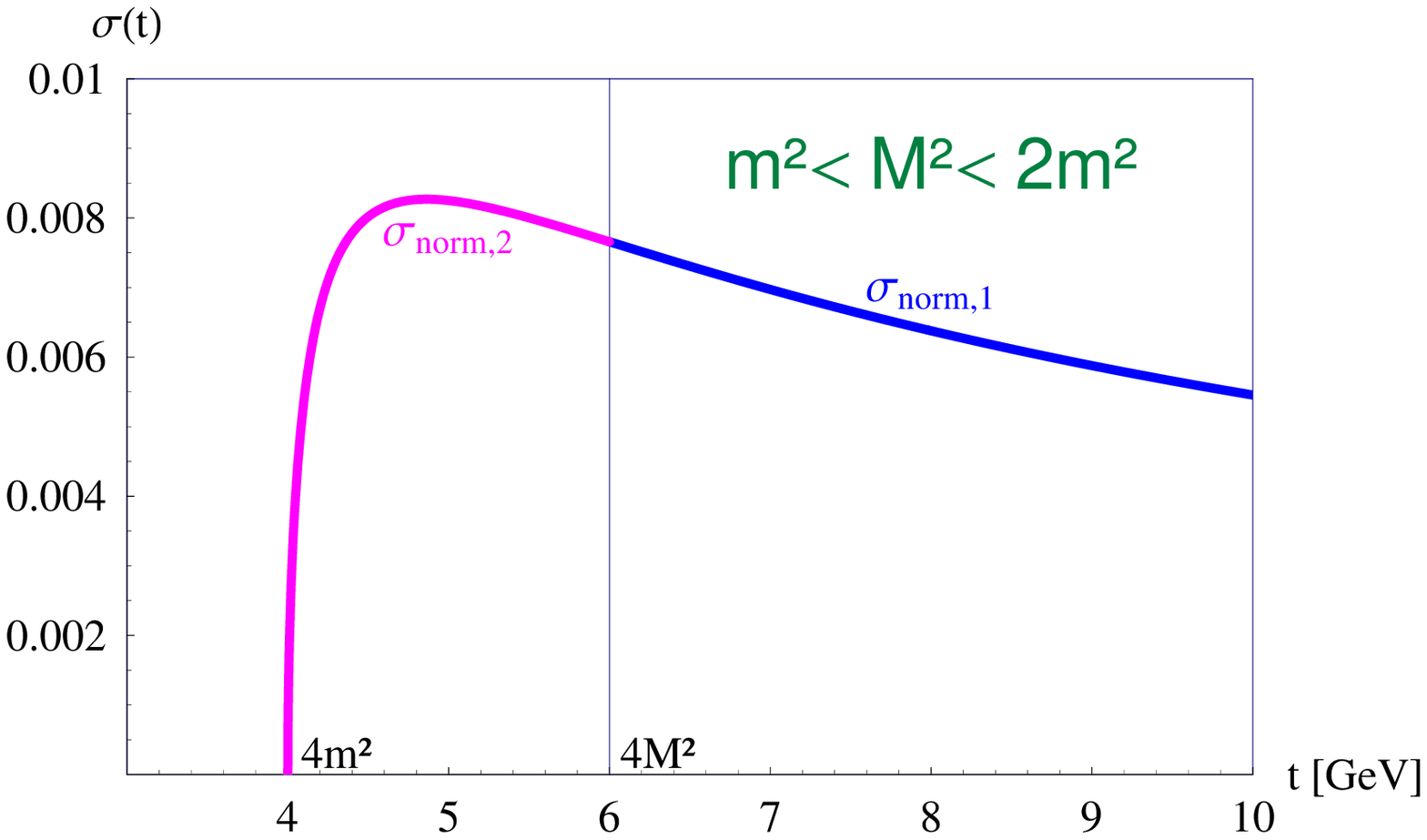}\\
\includegraphics[width=10.9cm]{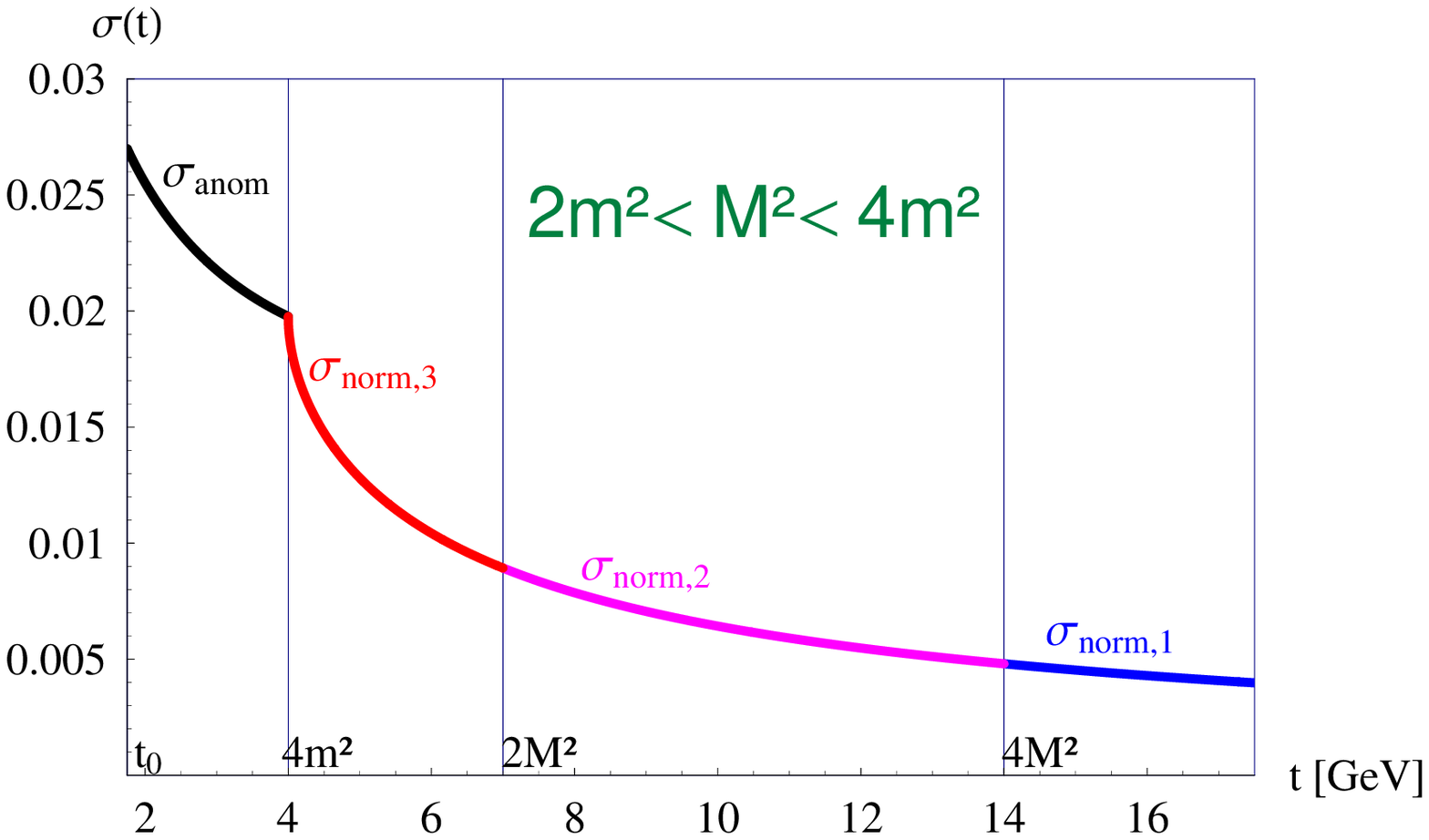}
\caption{\label{fig:1a}The spectral density $\sigma(s,M^2)$ for different values of $m$ and $M$. 
$\sigma_{\rm norm,1}$ denotes the expression in the first line of Eq.~(\ref{normM}); 
$\sigma_{\rm norm,3}$ denotes the expression in the second line of Eq.~(\ref{normM}); and 
$\sigma_{\rm norm,3}$ denotes the expression in the second line of Eq.~(\ref{normM}). 
$\sigma_{\rm anom}$ is given by Eq.~(\ref{anomM}). Notice that for $M^2<2m^2$, two upper plots, 
the spectral density vanishes at the normal threshold $t=4m^2$, whereas in the case $M^2>2m^2$, the spectral density does 
not vanish neither at the normal threshold $4m^2$, nor at the anomalous threshold $t_0$.}
\end{center}
\end{figure}

We start with calculating the spectral density in the region $M^2<0$, where it is given by Cutkosky rules:  
\begin{eqnarray}
\label{normMle0}
\sigma_{\rm norm}(t,M^2)=
\frac1{16\pi\sqrt{t(t-4M^2)}}
\log\left(
\frac{t-2M^2+\sqrt{t(t-4M^2)}\sqrt{1-4m^2/t}}{t-2M^2-\sqrt{t(t-4M^2)}\sqrt{1-4m^2/t}}
\right), \qquad M^2<0.
\end{eqnarray}
It turns out that this expression is also valid in a broader domain, namely, $M^2<m^2$ (see Fig.~\ref{fig:1a}a).  
Performing the analytic continuation in variable $M^2$ from the domain $M^2<m^2$, we obtain the spectral density 
for the region of interest $2m^2<M^2<4m^2$: 
\begin{eqnarray}
\label{normM}
\sigma_{\rm norm}(t,M^2)=
\left\{\begin{array}{ll} 
\frac1{16\pi\sqrt{t(t-4M^2)}}
\log\left(
\frac{t-2M^2+\sqrt{t(t-4M^2)}\sqrt{1-4m^2/t}}{t-2M^2-\sqrt{t(t-4M^2)}\sqrt{1-4m^2/t}}
\right), 
& \quad
4M^2 \le t,
\\ 
\frac1{8\pi
\sqrt{-t(t-4M^2)
}}\arctan
\left(
\frac{\sqrt{t(t-4M^2)}\sqrt{1-4m^2/t}}{t-2M^2}
\right), 
& 
\quad  
2M^2 \le t\le 4M^2,
\\ 
\frac1{8\pi
\sqrt{-t(t-4M^2)
}}
\left[\pi+\arctan\left(
\frac{\sqrt{t(t-4M^2)}\sqrt{1-4m^2/t}}{t-2M^2}
\right)\right], 
& 
\quad  4m^2\le t\le 2M^2.
\end{array}\right. 
\end{eqnarray}

Figure~\ref{fig:1a}~(b,c) shows the cases $m^2<M^2<2m^2$ and $2m^2<M^2<4m^2$, respectively. 
In the latter case, the normal spectral density does not vanish at the normal threshold $t=4m^2$. 
The discontinuity of the form factor $F(q^2,M^2)$ on the anomalous cut is related 
to the discontinuity of the function $\sigma_{\rm norm}(t,M^2)$ and reads 
\begin{eqnarray}
\label{anomM}
\sigma_{\rm anom}(t,M^2)=
\frac{1}{8\sqrt{-t(t-4M^2)}},\qquad t_0\le t\le 4m^2, \qquad t_0=\frac{M^2}{m^2}(4m^2-M^2).
\end{eqnarray} 
Therefore, the full spectral density has the form 
\begin{eqnarray}
\sigma(t,M^2)=
\theta(M^2-2m^2)
\theta(t_0\le t\le 4m^2)\sigma_{\rm anom}(t,M^2)
+
\theta(4m^2\le t)\sigma_{\rm norm}(t,M^2). 
\end{eqnarray}
Clearly, the spectral density given by Eqs.~(\ref{normM}) and (\ref{anomM}) is a continuous function for 
$t>t_0$ (see Fig.~\ref{fig:1a}~c). 

The spectral representation for the form factor reads  
\begin{eqnarray}
F(q^2,M^2)=
\theta(M^2-2m^2)
\int\limits_{t_0}^{4m^2}
\frac{dt}{\pi(t-q^2-i0)}\sigma_{\rm anom}(t,M^2)
+
\int\limits_{4m^2}^{\infty}
\frac{dt}{\pi(t-q^2-i0)}\sigma_{\rm norm}(t,M^2). 
\end{eqnarray}
It can be also written in the usual form 
\begin{eqnarray}
F(q^2,M^2)=\int\limits_{t_0}^{\infty}
\frac{dt}{\pi(t-q^2-i0)}\sigma(t,M^2),
\end{eqnarray}
where $\sigma(t,M^2)$ contains the normal and the anomalous pieces. 

\begin{figure}[!ht]
\begin{center}
\includegraphics[width=10cm]{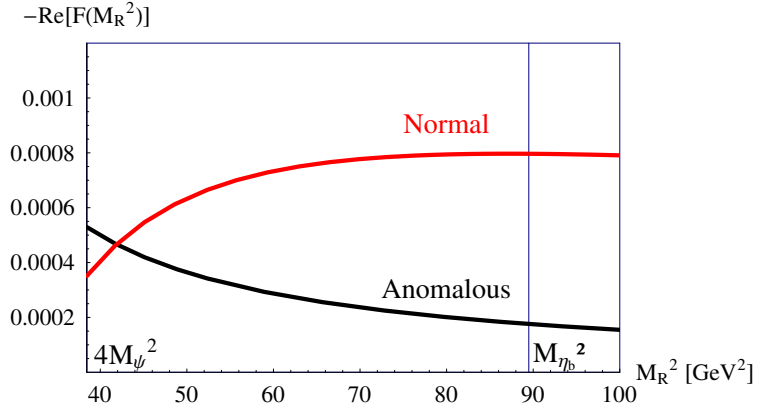}
\caption{\label{fig:1b}The real part of the normal (red) and of the anomalous (black) contributions to the 
amplitude $F(q^2=M_R^2,M^2)$ vs $M_R^2$. Here we use the following numbers $m=m_D=1.87$ GeV, $M=M_{J/\psi}=3.1$ GeV, 
and $M_{\eta_b}=9.46$ GeV.}
\end{center}
\end{figure}

Figure \ref{fig:1b} shows the calculated real parts of the normal and the anomalous contributions to the form factor 
$F(q^2=M_R^2,M^2)$ depending on the value of the mass of the decaying resonance $M_R^2=q^2>0$. 

%**************************************************************************
\newpage
\section{\label{sect:iii}
Double dispersion representation in $p_1^2$ and $p_2^2$ for particles of mass $m$ in the loop}
The previous Section presented the single dispersion representation in $q^2$ for the triangle diagram with 
particles of mass $m$ in the loop. For the same triangle diagram, a double dispersion representation 
in $p_1^2$ and $p_2^2$ may be written \cite{anisovich,anisovich_melikhov,lucha_melikhov}. 

At $q^2<0$, such a double dispersion representation in $p_1^2$ and $p_2^2$ has the form:  
\begin{eqnarray}
\label{double}
F(q^2,p_1^2,p_2^2)=\int \frac{ds_1}{\pi(s_1-p_1^2-i0)}\frac{ds_2}{\pi(s_2-p_2^2-i0)}
\Delta(q^2,s_1,s_2). 
\end{eqnarray} 
The double spectral density $\Delta(q^2,s_1,s_2)$ may be obtained by placing all three particles 
in the loop on the mass shell and taking the off-shell external momenta 
$p_1\to \tilde p_1$, $p_2\to \tilde p_2$, 
such that $\tilde p_1^2=s_1$, $\tilde p_2^2=s_2$, and 
$(\tilde p_1-\tilde p_2)^2=q^2$ is fixed \cite{anisovich}: 
\begin{eqnarray}
\Delta(q^2,s_1,s_2)&=&\frac1{8\pi}
\int dk_1 dk_2 dk_3 \delta(\tilde p_1-k_2-k_3)\delta(\tilde p_2-k_3-k_1)\nonumber\\
&&\times\theta(k_1^0)\delta(k^2_1-m^2)\theta(k_2^0)\delta(k^2_2-m^2)\theta(k_3^0)\delta(k^3_2-m^2), \nonumber \\
&&
\quad \tilde p_1^2=s_1, \quad \tilde p_2^2=s_2, \quad (\tilde p_1-\tilde p_2)^2=q^2. 
\end{eqnarray}
Explicitly, one finds 
\begin{eqnarray}
\label{thetafunction}
\Delta(q^2,s_1,s_2)&=&\frac{1}{16\lambda^{1/2}(s_1,s_2,q^2)}
\theta\left({s_1-4m^2}\right)
\theta\left({s_2-4m^2}\right)\nonumber\\
&&\times\theta\left[{\left(q^2(s_1+s_2-q^2)\right)^2-\lambda(s_1,s_2,q^2)\lambda(q^2,m^2,m^2)}\right]. 
\end{eqnarray}
The solution of the $\theta$-function gives the following allowed intervals 
for the integration variables $s_1$ and $s_2$: 
\begin{eqnarray}
\label{limits}
4m^2<&s_2,&
\nonumber\\
s_1^-(s_2,q^2)<&s_1&<s_1^+(s_2,q^2), 
\end{eqnarray}
where  
\begin{eqnarray}
\label{s1pm_equal_masses}
s_1^\pm(s_2,q^2)&=&s_2+q^2-\frac{s_2 q^2}{2 m^2}
\pm\frac{\sqrt{s_2(s_2-4m^2)}\sqrt{q^2(q^2-4m^2)}}{2m^2}.
\end{eqnarray}
The final double dispersion representation for the triangle diagram at $q^2<0$ takes the form\footnote{
The easiest way to obtain this double dispersion representation 
is to introduce light-cone variables in the Feynman expression, and to
choose the 
reference frame where $q_+=0$ (which restricts $q^2$ to $q^2<0$). Then the $k_-$ integral is easily done,
and the remaining $y$ and $k_\perp$ integrals may be written in the
form (\ref{double});  
details can be found in \cite{anisovich}.}
\begin{equation}
\label{fftrans}
F(q^2,p_1^2,p_2^2)=\int\limits^\infty_{4m^2}\frac{ds_2}{\pi(s_2-p_2^2-i0)}
\int\limits^{s_1^+(s_2,q^2)}_{s_1^-(s_2,q^2)}\frac{ds_1}{\pi(s_1-p_1^2-i0)}
\frac{1}{16\lambda^{1/2}(s_1,s_2,q^2)}.
\end{equation}
Notice the relation $s_1^-(s_2,q^2)>4m^2$, which holds for all $s_2>4m^2$ at $q^2<0$: this guarantees 
that the integration region in $s_1$ always remains above the normal threshold. 
Clearly, the integration region does not depend on the values of $p_1^2$ and $p_2^2$. 
Essential for us is that {\it no anomalous cuts emerge in the double dispersion 
representation in $p_1^2$ and $p_2^2$ for $q^2<0$}. This makes the double dispersion representation particularly 
convenient for treating the triangle diagram for values of $p_1^2$ and $p_2^2$ above 
the thresholds. 
One should just take care about the appearance of the absorptive parts. 

%\newpage
\section{\label{sect:iv}Double spectral representation for the decay kinematics}
Now we discuss the triangle diagram with particles of different masses in the loop, $m<\mu$, 
and consider the decay kinematics $0<q^2<(\mu-m)^2$ \cite{braun,melikhov}. 
We have in mind the application to processes corresponding to the overthreshold 
values $p_1^2>(\mu+m)^2$ and $p_2^2>4m^2$, such as, e.g., the $K\to 3\pi$ decay. 
As we have seen in the previous Section, the single dispersion representation in $q^2$ 
is rather complicated for $p_1^2$ and $p_2^2$ above the two-particle thresholds 
already for equal masses in the loop. 
The situation is much worse for unequal masses in the loop. 
On the other hand, we shall see that the double spectral representation in $p_1^2$ and 
$p_2^2$ is rather simple for $q^2<(\mu-m)^2$. 

We start with the region $q^2<0$, where the double dispersion representation has the 
standard form both for equal and unequal masses in the loop. 
We then perform the analytic continuation in
$q^2$ and observe the appearance of the anomalous contribution in the double spectral representation. 

\subsection{Transition form factor at $q^2<0$}
For $q^2<0$, the double dispersion representation has a form very similar to the case of 
equal masses \cite{melikhov}: 
\begin{equation}
\label{fftrans2}
F(q^2,p_1^2,p_2^2)=\int\limits^\infty_{4m^2}\frac{ds_2}{\pi(s_2-p_2^2)}
\int\limits^{s_1^+(s_2,q^2)}_{s_1^-(s_2,q^2)}\frac{ds_1}{\pi(s_1-p_1^2)}
\frac{1}{16\lambda^{1/2}(s_1,s_2,q^2)}, 
\end{equation}
where  
\begin{eqnarray}
\label{s1pm}
s_1^\pm(s_2,q^2)&=&
\frac{s_2(m^2+\mu^2-q^2)+2 m^2 q^2}{2m^2}
\pm\frac{\lambda^{1/2}(s_2,m^2,m^2)\lambda^{1/2}(q^2,\mu^2,m^2)}{2m^2}. 
\end{eqnarray}
A new feature compared with the case of equal masses in the loop  
is the appearance of the region $0<q^2<(\mu-m)^2$, which was absent in the equal-mass case.  
This region corresponds to the decay of a particle of mass $\mu$ to
a particle of mass $m$ with the emission of a particle of mass $\sqrt{q^2}$.

\subsection{Transition form factors at $q^2>0$}
The form factor in the region $0<q^2<(\mu-m)^2$ may be obtained by analytic 
continuation of the expression (\ref{fftrans}). 
Let us consider the structure of the singularities of the integrand in Eq.~(\ref{fftrans2}) 
in the complex $s_1$-plane for a fixed 
real value of $s_2$ in the interval $s_2>4m^2$. 
\begin{figure}
\begin{center}
\includegraphics[width=12cm]{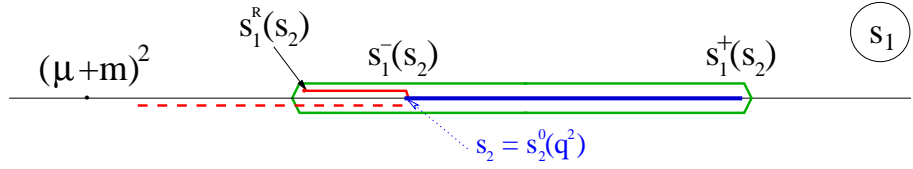}
\end{center}
\caption{\label{fig:double}
Singularities of the function $\Delta(q^2,s_1,s_2)$ in the complex $s_1$
plane as a function of $s_2$ for $q^2>0$. (This corresponds to the external $s_2$ 
and the internal $s_1$ integration). 
The trajectory $s_1^R(s_2)$ at fixed
$q^2>0$ is shown: for $s_2<s_2^0$ the branch point $s_1^R(s_2)$ 
remains on the unphysical sheet (dashed line), 
but, as soon as $s_2>s_2^0$, it goes onto the physical sheet and moves to the
left from the left boundary of the normal cut $s_1^-$.
Respectively, for $s_2<s_2^0$ the integration contour in the complex 
$s_1$-plane 
may be chosen along the interval $[s_1^-,s_1^+]$. For 
$s_2>s_2^0$, however, the contour should embrace the point $s_1^R$, and
therefore the integration contour contains two segments: the ``anomalous'' segment 
from $s_1^R$ to $s_1^-$, and the ``normal'' segment from $s_1^-$ to $s_1^+$.}
\end{figure}

\begin{figure}
\begin{center}
\includegraphics[width=12cm]{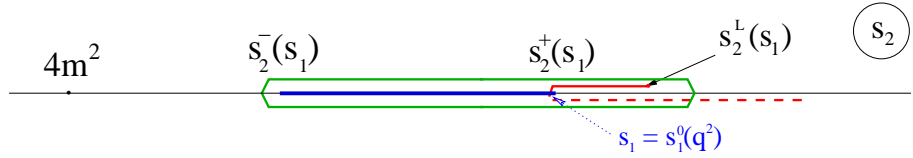}
\end{center}
\caption{\label{fig:double_b}
Singularities of the function $\Delta(q^2,s_1,s_2)$ in the complex $s_2$
plane as a function of $s_1$ for $q^2>0$. (This corresponds to the external $s_1$  
and the internal $s_2$ integration). 
The trajectory $s_2^L(s_1)$ at fixed
$q^2>0$ is shown: for $s_1<s_1^0$ the branch point $s_2^L(s_1)$ 
remains on the unphysical sheet (dashed line), 
but, as soon as $s_1>s_1^0$, it goes onto the physical sheet and moves to the
right from the right boundary of the normal cut $s_2^+$. The notations are self-evident. 
}
\end{figure}
The integrand has singularities (branch points) 
related to the zeros of the function $\lambda(s_1,s_2,q^2)$ at $s_1^L=(\sqrt{s_2}-\sqrt{q^2})^2$ and 
$s_1^R=(\sqrt{s_2}+\sqrt{q^2})^2$. As $q^2\le 0$, these singularities lie on the unphysical sheet. 
However, as $q^2$ becomes positive, the point $s_1^R$ may move onto the physical sheet through the cut 
from $s_1^-$ to $s_1^+$. This happens for values of the variable $s_2>s_2^0$, with 
$s_2^0$ obtained as the solution to the equation $s_1^R(s_2,q^2)=s_1^-(s_2,q^2)$. Explicitly, 
one finds  
\begin{eqnarray}
\sqrt{s_2^0}=\frac{\mu^2-m^2-q^2}{\sqrt{q^2}}.
\end{eqnarray}
The trajectory of the point $s_1^R(s_2,q^2)$ in the complex $s_1$-plane at fixed $q^2>0$ vs. $s_2$ 
is shown in
Fig.~\ref{fig:double}. 
%The situation is quite similar to the appearance of the anomalous cut in the single dispersion representation. 
As $q^2>0$, for $s_2>s_2^0(q^2)$ the
integration contour in the complex $s_1$-plane should be deformed such that it embraces the 
points $s_1^R$ and
$s_1^+$. Respectively, the $s_1$-integration contour contains the two segments: the normal part from 
$s_1^-$ to $s_1^+$, and the anomalous part from $s_1^R$ to $s_1^-$. The double spectral density 
for the anomalous piece is just the discontinuity of the function 
$1/\sqrt{\lambda(s_1,s_2,q^2)}$. It can be easily calculated as follows: 
Recall the relation $\sqrt{\lambda(s_1,s_2,q^2)}=\sqrt{s_1-s_1^L}\sqrt{s_1-s_1^R}$. The branch point 
$s_1^L$ lies on the unphysical sheet, therefore the function $\sqrt{s_1-s_1^L}$ is continuous 
on the anomalous cut located on the physical sheet. Thus we have to calculate the 
discontinuity of the
function $1/\sqrt{s_1-s_1^R}$ which is twice the function itself. As the result, the discontinuity
of the function $1/\sqrt{\lambda(s_1,s_2,q^2)}$ on the anomalous cut is $2/\sqrt{\lambda(s_1,s_2,q^2)}$.
Finally, the full double spectral density including the normal and the anomalous pieces 
takes the form
%\footnote{In \cite{braun,melikhov} the double spectral density was obtained by a rather
%complicated procedure, considering first the single spectral representation in $s_2$. We point out that
%this step is unnecessary: the final result may be obtained just starting from the double spectral 
%representation at $q^2<0$, where only the normal contribution is present. The derivation applied here 
%promises strong simplifications for obtaining the double spectral representation in the production region
%$q^2>(\mu + m)^2$.} 
\begin{eqnarray}
\label{4deltas}
\Delta(q^2,s_1,s_2|\mu,m,m)&=&
\frac{\theta(s_2-4m^2)\theta(s_1^-<s_1<s_1^+)}{16\lambda^{1/2}(s_1,s_2,q^2)}
+\frac{2\theta(q^2)\theta(s_2-s_2^0)\theta(s_1^R<s_1<s_1^-)}{16\lambda^{1/2}(s_1,s_2,q^2)}.
\end{eqnarray}
The first term in (\ref{4deltas}) relates to the Landau-type contribution
emerging when all
intermediate particles go on mass shell, while the second term describes the
anomalous contribution.

The result (\ref{4deltas}) for $\Delta$ holds for $\mu > m$ implying the ``external'' 
$s_2$-integration, and the ``internal'' $s_1$-integration. The location of the integration 
region for this case is shown in Fig.~\ref{fig:double}. Fig.~\ref{fig:double_b} gives the 
integration contour in the complex $s_2$ plane for the opposite integration order. 

\begin{figure}[h]
\begin{center}
\includegraphics[height=7.5cm]{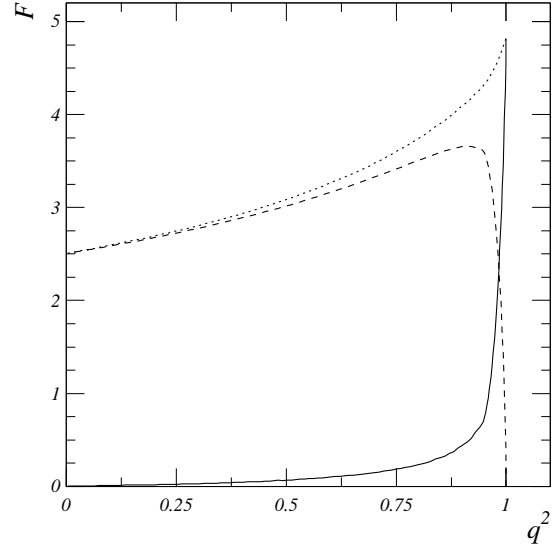}
\caption{\label{fig:plot} 
A typical behavior of the function 
$F(q^2,p_1^2,p_2^2)$ vs. $q^2$ for $0<q^2<(\mu-m)^2$ at
fixed $p_1^2$ and $p_2^2$. 
The parameters are chosen such that $(\mu-m)^2=1$ GeV$^2$. 
Dashed: normal part, solid: anomalous part, dotted: full function (sum of both parts). }
\end{center}
\end{figure}

\begin{figure}[h]
\begin{center}
\includegraphics[width=7cm]{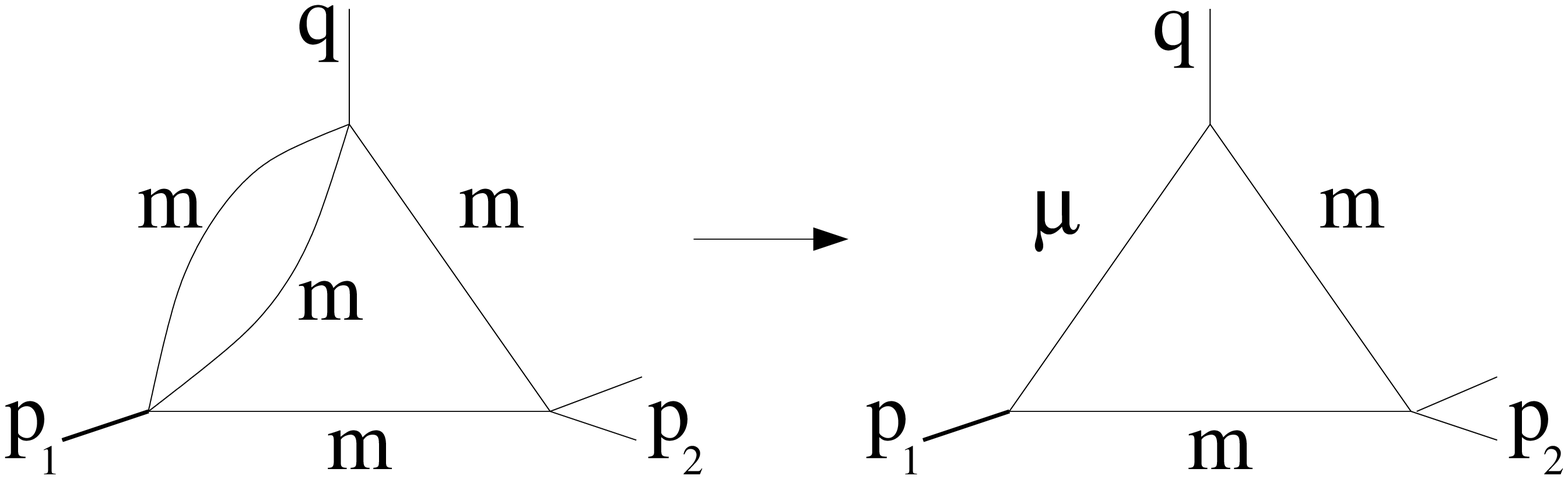}
\caption{\label{fig:k3pi} 
The triangle-diagram contribution to the $K\to 3\pi$ amplitude may be reduced to the 
integral over $\mu^2$ of the diagram $F$. }
\end{center}
\end{figure}

The final representation for the form factors at $0<q^2<(\mu-m)^2$
takes the form  
\begin{eqnarray}
\label{final1}
F(q^2,p_1^2,p_2^2)&=&
\int\limits_{4m^2}^\infty\frac{ds_2}{\pi(s_2-p_2^2-i0)}
\int\limits_{s_1^-(s_2,q^2)}^{s_1^+(s_2,q^2)}
\frac{ds_1}{\pi(s_1-p_1^2)}
\frac{1}{16\lambda^{1/2}(s_1,s_2,q^2)}
\\
&&+
2\theta\left(0<q^2<(\mu-m)^2\right)\int\limits_{s_2^0(q^2)}^\infty\frac{ds_2 }{\pi(s_2-p_2^2-i0)}
\int\limits_{s_1^R(s_2,q^2)}^{s_1^-(s_2,q^2)}
\frac{ds_1}{\pi(s_1-p_1^2)}\frac{1}{16\lambda^{1/2}(s_1,s_2,q^2)}.\nonumber
\end{eqnarray}
A typical behavior of the anomalous and the normal contributions is plotted in Fig.~\ref{fig:plot}:
the normal contribution first rises at small positive values of $q^2$ but then falls down steeply and 
vanishes at zero recoil. The anomalous contribution is zero at $q^2=0$, remains small at small 
$q^2>0$, but rises steeply near zero recoil, providing a smooth behavior of the full form factor. 
We point out that the representation (\ref{final1}) is particularly suitable for application to processes 
where $p_1^2$ and $p_2^2$ are above two-particle thresholds: in this case, the single spectral
representation in $q^2$ becomes extremely complicated, with a nontrivial integration contour in the
complex $q^2$-plane, whereas the double dispersion representation in $p_1^2$ and $p_2^2$ has the  
simple form given above. For values of $p_1^2$ and $p_2^2$ above the thresholds one just 
has to take into account the appearance of the absorptive parts in the $s_1$ and $s_2$ integrals. 
A possible application of this representation may be the calculation of the triangle-diagram
contribution to the three-body decay \cite{anisovich_anselm_1,anisovich_anselm}, e.g., 
to the $K\to 3\pi$ decay \cite{k3pi}, Fig.~\ref{fig:k3pi}. 
In the latter case, the diagram with the pion loop may be represented as the $\mu^2$ integral of the 
triangle diagram considered here, and one obtains the expression for the 
values $p_1^2=M_K^2>9m_\pi^2$, $p_2^2>4m_\pi^2$, and $q^2=m_\pi^2$. 
The emerging absorptive parts may then be easily calculated from the double spectral representation. 
The problem would be technically very involved if one uses the single spectral representation 
in $q^2$, as can be seen from the complicated structure of the integration contour in 
Section \ref{sect:ii}. 
 
%\newpage
\section{Summary and Conclusions} 
I have presented a detailed analysis of dispersion representations for the triangle diagram, laying main
emphasis on the appearance of the anomalous contributions in these representations. 
These anomalous singularities play an important role in the analysis of physical processes, see, e.g.,
\cite{santorelli,oka,mh,as,mm,guo2} and a recent review \cite{guo} for details. 
In some kinematic regions, the properties of the triangle diagram and the amplitudes of the corresponding processes are mainly determined by 
the anomalous contributions. A message I would like to convey to the reader
is that in many cases the double spectral representations in $p_1^2$ and $p_2^2$ provide great technical 
advantages compared to the use of the single representation in $q^2$. This is clearly the case for 
$p_1^2$ and $p_2^2$ above the thresholds and $q^2$ in the decay region $0<q^2<(\mu-m)^2$. 
Several realistic physical cases belong to this class of problems.   

Let me highlight a few points presented in this lecture: 

\begin{itemize}
\item
The single dispersion representation for the triangle diagram 
in the variable $q^2$ develops anomalous threshold, related to a migration of the logarithmic branch point from the unphysical 
(second) sheet of the Riemann surface onto the physical sheet through the normal $q^2$-cut. This migration occurs for the 
specific relationship between the squares of the external momenta $p_1^2$ and $p_2^2$ and the masses of the particles propagating 
in the loop. For instance, for the case $p_1^2=p_2^2=M^2$, and the particles with mass $m$ in the loop, the anomalous threshold 
occurs for all $M^2 > 2m^2$, i.e., for the external masses below the unitary thresholds $M^2=4m^2$. 
This means, in particular, that the anomalous thresholds occur for {\it weakly-bound} states, when $M-2m=-\epsilon_B$ and 
$\epsilon_B\ll m$. In this case, the location of the anomalous threshold in $q^2$ is responsible for the (large) radius of the 
weakly-bound state. 

It should be taken into account that the appearance of the anomalous cut in the single dispersion representation 
changes the leading singularity of the triangle diagram {\it on the physical sheet}: 

(a) In the ``normal'' case, $M^2 < 2m^2$, the spectral density of the triangle function has zero at the threshold, 
$\sigma(t)\sim \sqrt{t-4m^2}$ [Fig.~\ref{fig:1a}(a,b)], thus yielding the $\sqrt{q^2-4m^2}$ leading singularity 
of the triangle function. 

(b) In the ``anomalous'' case, $2m^2 < M^2$, the spectral density of the triangle function does not vanish at 
the anomalous threshold $t=t_0$  [Fig.~\ref{fig:1a}(c)], thus yielding the $\log(q^2-t_0)$ leading singularity. 
Emphasize that the logarithmic singularity does not emerge on the physical sheet in the ``normal'' case.  

\item
We pointed out that at spacelike momentum transfer $q$, $q^2<0$, and for any values of $p_1^2$ and $p_2^2$, 
the double dispersion representation in $p_1^2$ and $p_2^2$ is particularly simple and contains 
only the normal cuts. The calculation of the triangle diagram in this case may be easily done for all 
values of $p_1^2$ and $p_2^2$, including the values above the thresholds and complex values. In the same situation, 
the single spectral representation in $q^2$ contains, in addition, the anomalous cut, making the 
application of the single dispersion representation a very involved problem. 

\item
For the decay kinematics $0<q^2<(\mu-m)^2$, the anomalous thresholds and the anomalous cuts emerge in the double 
dispersion representations in the variables $p_1^2$ and $p_2^2$. In this case, the anomalous threshold in the 
variable $p_1^2$ (or $p_2^2$) is absent at $q^2<0$ but emerges for positive values of $q^2$. This anomalous threshold 
in this case lies above the normal unitary threshold at $p_1^2=(\mu+m)^2$ and $p_2^2=4m^2$.
The {\it anomalous} contribution is small at small positive $q^2$, but steeply rises 
when $q^2$ approaches the point $q^2=(\mu-m)^2$. On the contrary, the {\it normal} contribution dominates the 
form factor at small $q^2$, but vanishes at $q^2=(\mu-m)^2$. 

In the decay region $0<q^2<(\mu-m)^2$, the double spectral 
representation in $p_1^2$ and $p_2^2$ provides a very convenient  
tool for considering processes at $p_1^2$ and $p_2^2$ above the thresholds. 
The application of the single spectral representation in $q^2$ faces in this 
case severe technical problems. 
\end{itemize}

\acknowledgments
{I have great pleasure to express my gratitude to V.~Anisovich, W.~Lucha, H.~Sazdjian, and S.~Simula for 
collaboration on the problems discussed in this lecture. I would also like to thank the Organizers 
for creating a friendly and productive atmosphere of this School. 
}


\newpage

\begin{thebibliography}{99} 

\bibitem{karplus}
R.~Karplus, C.~Sommerfeld, and E.~Wichman, 
{\it Spectral Representations in Perturbation Theory. 
I. Vertex Functions}, 
Phys.~Rev.~{\bf 111}, 1187 (1958). 
\bibitem{landau}L.~D.~Landau, 
{\it On analytic properties of vertex parts in quantum field theory},
Nucl. Phys. {\bf 13}, 181 (1959).
\bibitem{fronsdal}
C.~Fronsdal and R.~E.~Norton, 
{\it Spectral Representations for Vertex Functions},
J.~Math.~Phys.~{\bf 5}, 100 (1964). 
\bibitem{norton}
R.~E.~Norton, {\it Location of Landau Singularities}, 
Phys.~Rev.~{\bf 6B}, 135, 1381 (1964).

\bibitem{burton}
G.~Burton, 
{\it Introduction to dispersion methods in field theory}, 
W.~A.~Benjamin, NY/Amsterdam, 1965. 
\bibitem{anisovich_book}
V.~V.~Anisovich, M.~N.~Kobrinsky, Y.~M.~Shabelsky, and J.~Nyiri, 
{\it Quark model and high-energy collisions}, World Scientific (2004). 
\bibitem{landau_lifshitz}
V.~B.~Berestetsky, E.~M.~Lifshitz, and L.~P.~Pitaevsky, 
{\it Quantum Electrodynamics}, 
Course of Theoretical Physics, vol. 4, Moscow, Nauka, 1989. 

\bibitem{iz}
C.~Itzykson and J.~B.~Zuber, 
{\it Quantum Field Theory}, 
International Series In Pure and
Applied Physics. McGraw-Hill, New York, 1980.
http://dx.doi.org/10.1063/1.2916419. 

\bibitem{zwicky}
R.~Zwicky, 
{\it A brief Introduction to Dispersion Relations 
and Analyticity}, arXiv:1610.06090. 

\bibitem{lms}
W.~Lucha, D.~Melikhov, and S.~Simula, 
{\it Dispersion representations and anomalous 
singularities of the triangle diagram}, 
Phys.~Rev.~{\bf D75}, 016001 (2007); 
Erratum: Phys.~Rev.~{\bf D92}, 019901 (2015).
%%%%%%%%%%%%%% Landau Eqs

\bibitem{hagop}
W.~Lucha, D.~Melikhov, and H.~Sazdjian, 
{\it Tetraquarks and two-meson states at large-$N_c$}, 
Eur.~Phys.~J.~{\bf C77}, 866 (2017);
{\it Tetraquark-adequate QCD sum rules for quark-exchange processes}, 
Phys.~Rev.~{\bf D100}, 074029 (2019). 

\bibitem{braun}
P.~Ball, V.~M.~Braun, H.~G.~Dosch, 
{\it Form-factors of semileptonic $D$ decays from QCD sum rules},
Phys.~ Rev.~{\bf D44}, 3567 (1991).

%%%%%%%%%%%%% 
\bibitem{procura}
G.~Colangelo, M.~Hoferichter, M.~Procura, and P.~Stoffer, 	
{\it Dispersion relation for hadronic light-by-light scattering: theoretical foundations}, 
JHEP {\bf 1509} 074 (2015).



% double for elastic
\bibitem{anisovich}
V.~V.~Anisovich, M.~N.~Kobrinsky, D.~I.~Melikhov, and A.~V.~Sarantsev, 
{\it Ward identities and sum rules for composite systems described in the dispersion relation 
technique: The Deuteron as a composite two nucleon system},  
Nucl.~Phys.~{\bf A544}, 747 (1992);\\  
V.~V.~Anisovich, D.~I.~Melikhov, B.~Ch.~Metsch, and H.~R.~Petry,
{\it The Bethe-Salpeter equation and the dispersion relation technique}, 
Nucl.~Phys.~{\bf A563}, 549 (1993). 
\bibitem{anisovich_melikhov}
V.~Anisovich, D.~Melikhov, V.~Nikonov, 
{\it Quark Structure of the Pion and Pion Form Factor},  
Phys.~Rev.~{\bf  D52}, 5295 (1995). 
V.~Anisovich, D.~Melikhov, V.~Nikonov,
{\it Photon-meson transition form factors 
$\gamma\pi$, $\gamma\eta$, and $\gamma\eta'$ 
at low and moderately high $Q^2$},  
Phys.~Rev.~{\bf D55}, 2918 (1997).
\bibitem{lucha_melikhov}
W.~Lucha and D.~Melikhov, 
{\it The $\gamma^*\gamma^*\to\eta_c$ transition form factor},
Phys.~Rev.~{\bf D86}, 016001 (2012).

% decay kinematics

\bibitem{melikhov}
D.~Melikhov, 
{\it Form-factors of meson decays in the relativistic constituent quark model}, 
Phys.~Rev.~{\bf D53}, 2460 (1996); 
D.~Melikhov, 	
{\it Heavy quark expansion and universal form-factors in quark model},  
Phys.~Rev.~{\bf D56}, 7089 (1997);
D.~Melikhov and B.~Stech, 
{\it Weak form-factors for heavy meson decays: An Update}, 
Phys.~Rev.~{\bf D62}, 014006 (2000);
D.~Melikhov,
{\it Dispersion approach to quark binding effects in weak decays of heavy mesons}, 
Eur.~Phys.~J.~direct {\bf 4} no.1, 2 (2002) [hep-ph/0110087]. 

% anomalous thresholds in three particles

\bibitem{anisovich_anselm_1}
V.~V.~Anisovich, A.~A.~Anselm, V.~N.~Gribov, and I.~T.~Dyatlov, 
{\it Anomalous thresholds and final state interaction}, 
Sov.~Phys.~JETP {\bf 16}, 643 (1963).
%Zh.Eksp.Teor.Fiz. 43 (1962) 906 
\bibitem{anisovich_anselm}
V.~V.~Anisovich and A.~A.~Anselm, 
{\it Theory of Reactions with Production of Three Partciles near Thresholds}, 
Sov.~Phys.~Uspekhi, {\bf 9}, 117 (1966). 
\bibitem{k3pi}
N.~Cabibbo and G.~Isidori,
{\it Pion-pion scattering and the $K\to 3\pi$ decay amplitudes}, 
JHEP {\bf 0503}, 021 (2005); \\ 
G.~Colangelo, J.~Gasser, B.~Kubis, and A.~Rusetsky, 
{\it Cusps in $K\to 3\pi$ decays}, 
Phys.~Lett.~{\bf B638}, 187 (2006). 

%%%%%%%%%%%%%%%%%%%% Applications in hadron physics
\bibitem{santorelli}
P.~Santorelli, {\it Long-distance contributions to the $\eta_b\to J/\psi J/\psi$ decay}, 
Phys.~Rev.~{\bf D77}, 074012 (2008).
\bibitem{oka}
X.-H.~Liu, M.~Oka, and Q.~Zhao, 
{\it Searching for observable effects induced by anomalous triangle singularities}, 
Phys.~Lett.~{\bf B753}, 297 (2016). 
\bibitem{mh}
M.~Hoferichter, B.-L.~Hoid, B.~Kubis, S.~Leupold, and S.~P.~Schneider,
{\it Pion-pole contribution to hadronic light-by-light scattering in the anomalous magnetic moment of the muon}, 
Phys.~Rev.~Lett.~{\bf 121}, 112002 (2018);
M.~Hoferichter, B.-L.~Hoid, B.~Kubis, S.~Leupold, and S.~P.~Schneider, 
{\it Dispersion relation for hadronic light-by-light scattering: pion pole}
JHEP {\bf 1810}, 141 (2018); 
M.~Hoferichter and P.~Stoffer, 
{\it Dispersion relations for $\gamma^*\gamma^*\to \pi\pi$: helicity amplitudes, subtractions, and anomalous thresholds}, 
JHEP~{\bf 1907}, 073 (2019). 
\bibitem{as}
A.~P.~Szczepaniak, 
{\it Triangle Singularities and $XYZ$ Quarkonium Peaks}, 
Phys.~Lett. {\bf B747}, 410 (2015); 
{\it Dalitz plot distributions in presence of triangle singularities},  
Phys.~Lett.~{\bf B757}, 61 (2016). 
\bibitem{mm}
M.~Mikhasenko, 
{\it A triangle singularity and the LHCb pentaquarks}, arXiv:1507.06552 [hep-ph];
M.~Mikhasenko, B.~Ketzer, and A.~Sarantsev, {\it Nature of the $a_1(1420)$}, 
Phys.~Rev.~{\bf D91}, 094015 (2015). 
\bibitem{guo2}
F.-K.~Guo, U.-G.~Mei{\ss}ner, W.~Wang, and Z.~Yang, 
{\it How to reveal the exotic nature of the $P_c(4450)$}, 
Phys.~Rev.~{\bf D92}, 071502 (2015);  
X.-H.~Liu, Q.~Wang, and Q.~Zhao, 
{\it Understanding the newly observed heavy pentaquark candidates}, 
Phys.~Lett.~{\bf B757}, 231 (2016);  
R.~Aaij (LHCb Collaboration), 
{\it Observation of a narrow pentaquark state, $P_c(4312)^+$, and of two-peak structure of the $P_c(4450)^+$}
Phys.~Rev.~Lett.~{\bf  122}, 222001 (2019). 
\bibitem{guo}
F.-K.~Guo, X.-H.~Liu, and S.~Sakai,  
{\it Threshold cusps and triangle singularities in hadronic reactions}, 
arXiv:1912.07030 [hep-ph]. 
\end{thebibliography}
\end{document}